\documentclass[%
 aip,
 jmp,%
 amsmath,amssymb,
reprint,%
]{revtex4-1}

\usepackage{graphicx}
\usepackage{dcolumn}
\usepackage{bm}
\usepackage{braket}
\usepackage{bbold}

\begin{document}

\preprint{APS/123-QED}

\title{Universal finestructure eraser for quantum dots}

\author{A. Fognini}
\email{a.w.fognini@tudelft.nl}
\affiliation{
Kavli Institute of Nanoscience Delft,\\
Delft University of Technology, Delft 2628 CJ, The Netherlands
}%

\author{A. Ahmadi}
 \affiliation{%
  Institute for Quantum Computing and Department of Physics \& Astronomy, \\
  University of Waterloo, Waterloo, ON N2L 3G1, Canada
}%
\author{S. J. Daley}
\author{M. E. Reimer}
 \affiliation{%
  Institute for Quantum Computing and Department of Electrical \& Computer Engineering, \\
  University of Waterloo, Waterloo, ON N2L 3G1, Canada
}%
\author{V. Zwiller}%
 \affiliation{%
  Department of Applied Physics, Royal Institute of Technology (KTH), \\
  AlbaNova University Center, SE - 106 91 Stockholm, Sweden
}%
\affiliation{
Kavli Institute of Nanoscience Delft,\\
Delft University of Technology, Delft 2628 CJ, The Netherlands
}%

\altaffiliation{Kavli Institute of Nanoscience Delft,\\
Delft University of Technology, Delft 2628 CJ, The Netherlands}


\date{\today}

\begin{abstract}
The measurable degree of entanglement from a quantum dot via the biexciton-exciton cascade depends crucially on the bright exciton fine-structure splitting and on the detection time resolution. Here, we propose an optical approach with fast rotating waveplates to erase this fine-structure splitting and therefore obtain a high degree of entanglement with near-unity efficiency. Our optical approach is possible with current technology and is also compatible with any quantum dot showing fine-structure splitting.
\end{abstract}

\pacs{78.67.Hc, 78.67.Uh, 73.21.La, 78.60.Hk, 78.47.jd, 42.50.Ex}
\keywords{}
\maketitle


Semiconductor quantum dots (QDs) allow for the generation of polarization entangled photons \cite{PhysRevLett.84.2513, salterNat, hafenbrak} through the biexciton-exciton cascade \cite{PhysRevLett.87.183601}.
Effects such as QD shape elongation \cite{PhysRevLett.95.257402, PhysRevLett.103.063601}, piezoelectric fields \cite{PhysRevLett.95.257402}, inhomogeneous alloy composition \cite{doi:10.1021/nl503581d, PhysRevLett.103.063601}, strain fields \cite{doi:10.1063/1.2204843}, or more generally all effects lowering the symmetry of the exciton's trapping potential \cite{PhysRevLett.103.063601} lead to a splitting of the exciton state.
The spin-degeneracy of the bright exciton level is therefore normally split in QDs due to the spin-orbit interaction \cite{PhysRevB.67.161306}. This splitting is called the fine-structure splitting (FSS) and its energy scale typically lies between $0-100\,\mu$eV in the case of III-V semiconductor quantum dots \cite{doi:10.1021/nl503581d}. The FSS introduces a which-path information during the biexciton-exciton decay whereby it was argued as being one of the main reasons for lowering the polarization entanglement \cite{PhysRevB.67.085317, stevensonNat}. QD growth methods have been successfully developed to minimize the FSS \cite{PhysRevB.88.041306, JGediminas, NanowireEntanglement}, but reaching vanishing FSS remains a significant challenge.
Consequently, several post-growth techniques have been developed to solve this problem by tuning the FSS to zero. Compensation has been achieved through external strain fields \cite{doi:10.1063/1.2204843, PhysRevLett.114.150502}, magnetic fields \cite{stevensonNat}, electric fields \cite{doi:10.1063/1.1855409, PhysRevLett.103.217402}, annealing \cite{PhysRevB.69.161301}, or a combination of these approaches \cite{PhysRevLett.109.147401}. 
Typically, these techniques act macroscopically on the sample and only fully compensate \emph{one} out of millions of QDs. Scaling up to many quantum dots on the same sample is a challenge.
Furthermore, the well established strain compensation technique \cite{doi:10.1063/1.2204843, PhysRevLett.114.150502} is difficult to adapt for QDs embedded in photonic nanostructures \cite{NanowireEntanglement, doi:10.1021/nl503581d, SomaschiNat, PhysRevB.90.201408,PhysRevLett.110.177402} due to strain relaxation over a length scale of $\approx100\,\mathrm{nm}$ \cite{PhysRevB.90.201408}.
Quantum dots embedded in nanowires \cite{NanowireEntanglement, PhysRevLett.110.177402, PhysRevB.90.201408, doi:10.1021/nl503581d} and micropillar cavities \cite{gazzanoNatCom, SomaschiNat}; however, possess several benefits such as enhanced photon extraction due to directional emission and near-unity single mode fiber coupling \cite{bulgariniNanoLett, Wang2017}. Therefore, a universal FSS compensation technique easily applicable to QDs would be of great value.

In this paper we introduce a novel FSS universal eraser technique to enhance the measurable entanglement towards unity by using frequency shifting capabilities of rotating $\lambda/2$-waveplates applied to both X and XX photons. Of particular significance, this frequency conversion process occurs without loss of photons due to only unitary optical manipulations. Our approach differs from Ref. \cite{doi:10.1063/1.3427485} as it can be implemented with current technology, 
is not intrinsically slow ($\approx10\,\mathrm{kHz}$) due to high voltage sweeps, does not rely on the splitting of different polarization modes \cite{PhysRevA.73.033813}, and requires only half of the frequency as compared to Ref. \cite{PhysRevA.73.033813} to compensate the FSS.

\begin{figure}
    \includegraphics{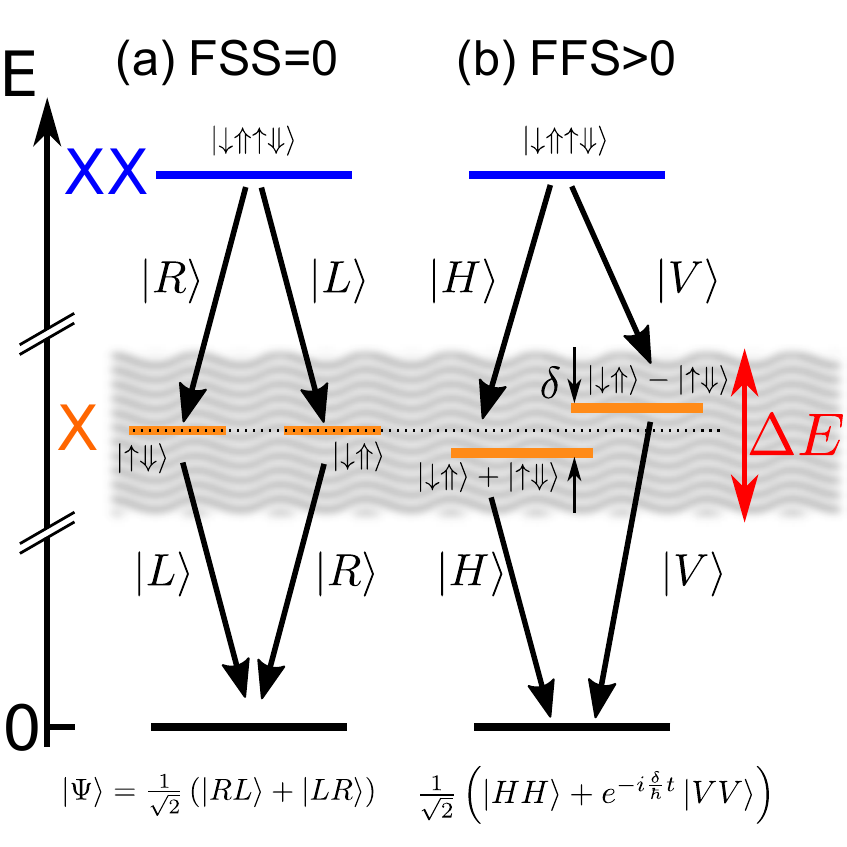}
    \caption{Representation of the biexciton (XX) exciton (X) emission. (a) In case of zero FSS the X-levels are degenerate and the two decay paths are indistinguishable which creates the entangled photon state $\frac{1}{\sqrt{2}}\left(\Ket{RL}+\Ket{LR}\right)$. 
    (b) For non-zero FSS the X-level is split by $\delta$ and the quantum state will precess between these two states. 
    However, with a fast measurement ($\Delta E \geq \delta$) the two X states (in H/V basis) cannot be resolved anymore which entangles the photons. The uncertainty introduced through the measurement process is indicated by the wavy gray background.}
    \label{fig:cascade}
\end{figure}

We first discuss the influence of the detection system's time resolution on the measurable entanglement. The term detection system includes every component used to detect the arrival time of the two photons from the cascade, e.g., detector time jitter, the electronics to correlate the arrival times of the biexciton and exciton photons, and dispersion in optical components. We define the full width at half maximum of the correlation time distribution of such a system as the time resolution $\tau$. For the sake of clarity we only consider FSS for reducing the measurable entanglement by phase averaging and do not consider dephasing mechanisms \cite{PhysRevLett.99.266802}.
Fig. \ref{fig:cascade} (a) depicts the biexciton-exciton cascade without FSS.
The cascade starts by the radiative decay of the biexciton (XX) state. Either a right- or left polarized \emph{single} photon is emitted ($\Ket{R}$, $\Ket{L}$) \cite{PhysRevLett.95.257402}. After the emission of the XX photon the system is in the exciton state (X). 
This level is degenerate and $\Ket{\uparrow\Downarrow}$, $\Ket{\downarrow\Uparrow}$ are the state's eigenfunctions in spin space \cite{PhysRevLett.95.257402}. 
Here, $\uparrow, \downarrow$ and $\Uparrow, \Downarrow$ denote the electron and hole spins, respectively. Since we assumed zero FSS, it is impossible to know whether a spin up or down electron has recombined. This lack of knowledge entangles the photons to $\Ket{\Phi} = \frac{1}{\sqrt{2}}\left(\Ket{RL}+\Ket{LR}\right)$. In this situation, the detection system's time resolution does not affect the measurable entanglement of this state since it does not change over time.

The situation is quite different in the case of finite FSS, as illustrated in \mbox{Fig. \ref{fig:cascade} (b)}. Due to spin-orbit interaction the exciton states mix and the new eigenfunctions become $\frac{1}{\sqrt{2}}\left(\Ket{\downarrow\Uparrow}-\Ket{\uparrow\Downarrow}\right)$ and $\frac{1}{\sqrt{2}}\left(\Ket{\downarrow\Uparrow}+\Ket{\uparrow\Downarrow}\right)$ \cite{Poem2010}.
After the XX decay the X will precess between these two eigenfunctions until it also decays. This evolution makes the quantum state time dependent \cite{PhysRevLett.101.170501} and reads as
\begin{equation}
\Ket{\Psi(t,\delta)} = \frac{1}{\sqrt{2}}\left(\Ket{HH}+e^{-i \frac{\delta}{\hbar}t}\Ket{VV}\right),
\label{equ:timeevol}
\end{equation}
where $\delta$ is the FSS energy, and $\Ket{H}$ and $\Ket{V}$ denote horizontally and vertically polarized single photon states. Equation \ref{equ:timeevol} describes a fully entangled state even with finite FSS as shown by Stevenson et al. \cite{PhysRevLett.101.170501}.
In fact, a slow detection system ($\tau\gg\hbar/\delta$) will average out the exponential phase term \cite{PhysRevLett.101.170501} in equation \ref{equ:timeevol} and only classical correlations are detected \cite{PhysRevB.66.045308}.
In contrast, a fast detection system ($\tau\ll\hbar/\delta$) will render the two decay pathways indistinguishable since the energy uncertainty relation $\Delta E \geq \frac{\hbar}{2\tau}$ does not allow for a precise energy measurement anymore. This point of view is complementary to spectral filtering \cite{Akopian2006, doi:10.1063/1.2722769} where only states with the same energy are analyzed but at the expense of filtering off many entangled photons.
Several experiments \cite{doi:10.1021/nl503581d, PhysRevLett.101.170501} have been performed making use of this uncertainty effect resolving the time evolution of the photonic state in equation \ref{equ:timeevol}. 
Nevertheless, a finite detector time resolution \emph{always} introduces phase averaging and inevitably reduces the measurable entanglement.

\begin{figure}
    \includegraphics[width=86mm]{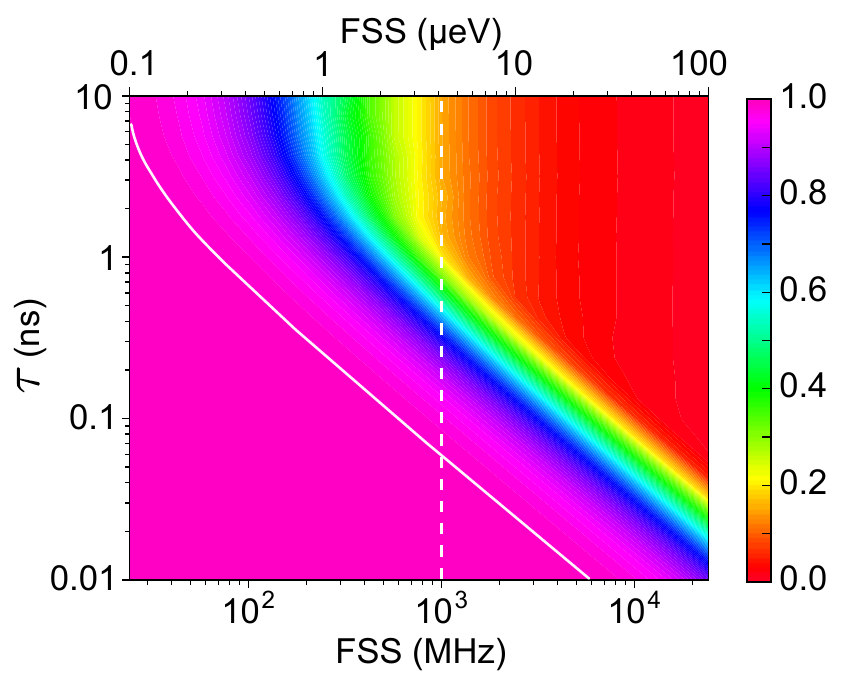}
    \caption{The measurable entanglement represented as the averaged concurrence $\bar{\mathcal{C}}$ as a function of the detector time resolution ($\tau$) and fine-structure splitting (FSS) in case of an exciton lifetime of $\tau_X=1\,\mathrm{ns}$. The white dashed line is a guide to the eye for the examples in the text and the white solid line highlights the $0.99$ contour line.}
    \label{fig:simu}
\end{figure}

In the following we will quantify this effect. In a quantum state tomography measurement \cite{PhysRevA.64.052312} the state described in equation \ref{equ:timeevol} is projected on the measurement basis $\Bra{ij}$, where $\mathrm{i,j} \in \Set{H,V,D,A,R,L}$ with $D$, $A$ denoting the diagonal and antidiagonal polarization states, respectively.
We define the time evolution of the measured biexciton-exciton pair rate as $n(t,\tau_X)=\frac{N_0}{\tau_X} e^{-t/\tau_X}$ for $t\geq0$ and $n(t,\tau_X)=0$ otherwise. Here, $t$ denotes the time after biexciton emission, $\tau_X$ the lifetime of the exciton level, and $N_0$ the number of detected photon pairs.
In case of perfect time resolution we get a time dependent correlation rate in each projection ${i,j}$ as
\begin{equation}
n_{i,j}(t,\delta,\tau_X)={|\Braket{ij|\Psi(t,\delta)}|}^2n(t,\tau_X).
\label{equ:nij}
\end{equation}
The effect of finite time resolution of the detection system is modeled by $g(t,\tau)$, a Gaussian with full width at half maximum of $\tau$. In such circumstances the detected projections are given by a convolution of the detection time resolution with equation \ref{equ:nij} yielding
\begin{equation}
m_{i,j}(t,\delta,\tau,\tau_X) = n_{i,j}(t,\delta,\tau_X)*g(t,\tau).
\end{equation}
The amount of entanglement which remains in $m_{i,j}(t,\delta,\tau,\tau_X)$ can be quantified by its concurrence $\mathcal{C}$ which lies between zero and one. One in case the system is fully entangled and zero if there are only classical correlations present. Since the state with finite FSS is evolving in time we define the time averaged concurrence $\bar{\mathcal{C}}$ weighted with the amount of detected photons per infinitesimal time bin as
\begin{equation}
\bar{\mathcal{C}}(\delta,\tau, \tau_X) := \lim_{T\rightarrow\infty}\frac{1}{N_0}\int_{-T}^{T}n(t)\mathcal{C}\left(\rho\left(m_{i,j}\right)\right)dt,
\label{equ:chat}
\end{equation}
where $\rho(m_{i,j})$ denotes the density matrix reconstructed from $m_{i,j}(t,\delta,\tau,\tau_X)$. Equation \ref{equ:chat} is evaluated numerically \footnote{Tomography library: https://github.com/afognini/Tomography} for an exciton lifetime of $\tau_X=1\,\mathrm{ns}$ and the result is presented in Fig. \ref{fig:simu}.
The result indicates that with sufficiently fast detection, perfect entanglement can be reconstructed. With a state of the art detection system a time resolution of $\tau=20\,\mathrm{ps}$ is possible. A FSS of $\delta=1\,\mathrm{GHz}$ (white dashed line in Fig. \ref{fig:simu}), yields a measurement of $\bar{\mathcal{C}}=0.999$ very close to unity. With regular avalanche photodiodes of $\tau=300\,\mathrm{ps}$ this value already reduces to $\bar{\mathcal{C}}=0.77$. Worsening the detection system resolution further to a time resolution of $\tau=1\,\mathrm{ns}$ yields almost no entanglement. In this latter case, the concurrence significantly reduces to $\bar{\mathcal{C}}=0.19$.
However, the latter nanosecond time resolution would be preferred in applications regarding secure communication protocols where accurate timing on picoseconds over kilometers \cite{ursinNatPhys} becomes difficult.

In the following we introduce a method to compensate FSS, making it possible to employ detection systems with any timing resolution smaller than the QD's photon repetition period such that no overlap between adjacent pulses occur.
The evolution of equation \ref{equ:timeevol} with finite FSS is unitary due to the time evolving exponential phase factor. Thus, it must be possible \cite{PhysRevLett.101.170501} to undo this phase evolution by suitable unitary optical components. 
The main component to achieve complete removal is a rotating $\lambda/2$-waveplate. Such an optical component acts on circularly polarized light 
as a single sideband frequency shifter \cite{Page1970, 555095}. A $\lambda/2$-waveplate spinning with angular frequency $\omega$ acting on a photon state can thus be described by the following operator
\begin{equation}
\Lambda_{1/2}\left(\omega\right)=\sum_{k}{a_{k+\frac{2\omega}{c},L}^\dagger a_{k,R}+a_{k-\frac{2\omega}{c},R}^\dagger a_{k,L}},
\label{equ:shifter}
\end{equation}
\noindent
where only k-vectors ($k$) perpendicular to the plane of the waveplate are considered.
Here, $c$ denotes the speed of light, and $a_{k,\lambda}$, $a_{k,\lambda}^{\dagger}$ denote anhilation and creation operators of photons with wavevector of length $k$ and right or left circular polarization $\lambda\in \{R,L\}$, respectively.
The action of a rotating $\lambda/2$-waveplate as described by equation \ref{equ:shifter} will up-convert $\Ket{R}$ photons by the energy $2\hbar\omega$ and down-convert $\Ket{L}$ photons by the same amount. Remarkably, this process can be achieved with unity efficiency.
\begin{figure}
    \includegraphics{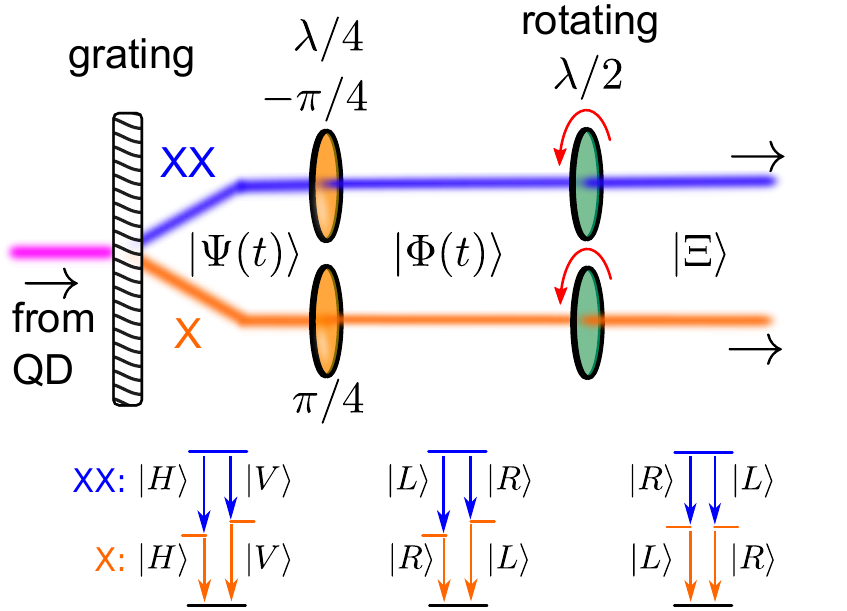}
    \caption{Proposed optical setup to compensate for a finite FSS. First, a polarization insensitive transmission grating splits the biexciton (XX) from the exciton (X) line. Next, a $\lambda/4$-plate transforms the X and XX photons into the circular basis. Finally, a $\lambda/2$-plate (one for each photon) rotating with an angular frequency of $f=\frac{\delta}{8\pi\hbar}$ compensates for the FSS. The polarization of the photons are indicated underneath the optical path after each waveplate. The length of the arrows is indicative for the photon energy. For convenience possible mirrors have been omitted.
    }
    \label{fig:setup}
\end{figure}
With the help of two $\lambda/4$-waveplates the XX and the X state transform into
\begin{equation}
\begin{split}
\Ket{\Phi(t,\delta)}&=\Lambda_{1/4}(-\pi/4)\otimes\Lambda_{1/4}(\pi/4)\Ket{\Psi(t,\delta)}\\
&=\frac{1}{\sqrt{2}}\left(\Ket{LR}+e^{-i \frac{\delta}{\hbar}t}\Ket{RL}\right),
\end{split}
\label{equ:trans}
\end{equation}
where the angles $\pm\pi/4$ are oriented with respect to the horizontal orientation. 
Now, sending this new state $\Ket{\Phi(t,\delta)}$ through a spinning $\lambda/2$-waveplate rotating with angular frequency of $\omega=\frac{\delta}{4\hbar}$ yields an entangled Bell state
\begin{equation}
\begin{split}
\Ket{\Xi}&=\Lambda_{1/2}(t,\omega)\otimes\Lambda_{1/2}(t,\omega)\Ket{\Phi(t,\delta)} \\
&= \frac{1}{\sqrt{2}}\left(\Ket{RL}+\Ket{LR}\right),
\end{split}
\end{equation}
where the time dependent phase factor has been completely removed. Here, $\Lambda_{1/2}(t,\omega)$ represents the operator from equation \ref{equ:shifter} in the time domain.
A setup to erase the FSS is depicted in Fig. \ref{fig:setup}.
First, a dispersive element, such as a high efficiency transmission grating, splits the XX line from the X line. Next, the XX (X) photon is sent through a fixed $\lambda/4$ waveplate offset from the horizontal direction by $-\pi/4$ ($\pi/4$). The photon state at this stage is represented by equation \ref{equ:trans}.
Finally, letting them both pass through a rotating $\lambda/2$-waveplate with angular frequency $\omega=\frac{\delta}{4\hbar}$ removes the FSS completely. For a possible implementation of a rotating $\lambda/2$-waveplate, electro-optical modulators (EOM) can be employed \cite{Qin:17, 4257052}. 
Here, the conversion efficiency is only limited by the transmission through the EOM. In work on EOMs, due to the high transparency a $95\,\%$ conversion efficiency was achieved \cite{Noe}.

An additional benefit of this approach is that the RF frequency is relatively low due to the fact that a rotating $\lambda/2$-waveplate shifts the frequency by twice its modulation frequency. For example, the RF frequency ($f=\frac{\omega}{2\pi}=\frac{\delta}{8\pi\hbar}$) necessary to compensate a FSS of $10\,\mu\mathrm{eV}$ is $604.5\,\mathrm{MHz}$, which should be easily achievable with current EOM technology reaching tens of GHz modulation bandwidth.
As the proposed technique is not invasive on the sample containing the QDs it is possible to compensate for the FSS of all QDs on the same sample as the FSS eraser technique can be done after propagation before detection.

In conclusion, we have analyzed the effect of finite FSS and the influence of the detection time resolution on the measurable entanglement from a single QD via the biexciton-exciton cascade.
The uncertainty in energy and time in the measurement allows the emitted QD photons to be entangled when a detection system with sufficient timing resolution is employed. However, the precise timing requirement on a picosecond level is hampering the progress in making the entanglement useful for applications and research. We have proposed a universal optical setup to completely remove the FSS based on a rotating $\lambda/2$-plate, which can be implemented with current EOM technology. 
The proposed technique will allow to make the entanglement created from QDs available for many applications like quantum communication, sensing, and imaging. Furthermore, it allows to compensate the FSS from many QDs from the same sample as macroscopic fields acting on a single QD are completely avoided.

A. Fognini gratefully acknowledges the Swiss National Science Foundation for the support through their Early PostDoc Mobility Program and thanks I. Esmaeil Zadeh for discussing the topic. M.E. Reimer acknowledges Industry Canada and NSERC for support.

%

\end{document}